\documentclass[journal]{IEEEtran}

\usepackage{cite}
\usepackage{float}
\usepackage{amsmath}
\usepackage{amsfonts}
\usepackage{graphicx}  
\usepackage{textcomp}    
\usepackage{framed}
\usepackage{algorithm,algpseudocode}

\usepackage{mathtools}
\DeclarePairedDelimiter{\ceil}{\lceil}{\rceil}

\ifCLASSINFOpdf
\else
\fi
\begin{document}
%
\title{Subspace Identification of Temperature Dynamics}
%
%
%

\author{Aleksandar~Haber 
\thanks{A. Haber is with the Department
of Engineering and Environmental Science, The City University of New York, College of Staten Island, New York,
NY, 10304 USA, e-mail: aleksandar.haber@csi.cuny.edu.}}
\maketitle

\begin{abstract}
 Data-driven modeling and control of temperature dynamics in mechatronics systems and industrial processes are challenging control engineering problems. This is mainly because the temperature dynamics is inherently infinite-dimensional, nonlinear, spatially distributed, and coupled with other physical processes. Furthermore, the dominant time constants are usually long, implying that in practice due to various economic and time constraints,  we can only collect a relatively small number of data samples that can be used for data-driven modeling. Finally, since sensing and actuation of temperature dynamics are often spatially discrete, special attention needs to be given to sensor (actuator) placement and identifiability problems. Motivated by these challenges, in this manuscript, we consider the problem of data-driven modeling and validation of temperature dynamics. We have developed an experimental setup consisting of a long aluminum bar whose temperature dynamics is influenced by spatially distributed heat actuators and whose temperature is sensed by spatially distributed thermocouples. We address the noise reduction problem and perform step response and nonlinearity analyses. We combine predictor based subspace identification methods with time series analysis methods to identify a multiple-input multiple-output system model. We provide detailed treatments of model structure selection, validation, and residual analysis problems under different modeling and prediction scenarios. Our extensive experimental results show that the temperature dynamics of the experimental setup can be relatively accurately estimated by low-order models.
\end{abstract}


%
\IEEEpeerreviewmaketitle

\section{INTRODUCTION}

 Data-driven modeling, estimation, and control of temperature dynamics are ubiquitous control engineering problems appearing in a number of mechatronics systems, industrial applications, and scientific fields. For example, heating and thermally induced deformations of lenses and mirrors in optical lithography machines used for integrated circuit manufacturing can seriously degrade the quality of manufactured circuits~\cite{bakshi2009euv}. To develop methods and actuators to counteract these thermally induced deformations, it is of paramount importance that the temperature dynamics of optical elements is accurately modeled and controlled~\cite{haber2013predictive,haber2013identification,saathof2016deformation,polo2013linear,ravensbergen2013deformable,haber2013iterative,haber2018sparsity}. Besides this,  modeling and control of temperature dynamics are important for temperature estimation and health monitoring of lithium-ion batteries~\cite{lin2013online}, as well as for temperature control of  greenhouses~\cite{hidayat2017identification,nielsen1998identification}, buildings~\cite{ma2012model,ma2012predictive,genc2017parametric}, and for proper operation of a number of industrial processes and systems~\cite{seborg2010process,kasprzack2013performance}.
 
In practice, it is often the case that we do not know \textit{a priori} all the parameters, equations, and boundary conditions that mathematically describe the dynamics of a process. Consequently, to obtain sufficiently accurate models that can be used for model-based observer and control designs, we need to employ data-driven model estimation methods, such as system identification techniques~\cite{ljung1998system,verhaegen2007filtering}. However, in practice, the system identification of the temperature dynamics might be a challenging problem due to the following reasons.  Namely, in its essence, the temperature dynamics is governed by a heat equation that is coupled with linear convection and nonlinear radiation boundary conditions. That is, the temperature dynamics is inherently infinite-dimensional. Furthermore, in many engineering applications, it is coupled with other physical processes such as elastic deformations and a fluid flow~\cite{incropera1985introduction,myers1985development,kasprzack2013performance}. Consequently, the structure selection and model order estimation of the temperature dynamics are challenging system identification problems.   Furthermore, time constants of the temperature dynamics are usually long,  implying that in practice due to various economic and time constraints,  we can only collect a relatively small number of data samples that can be used for system identification. Furthermore, the dynamics is usually actuated and sensed at a small number of spatial locations. Consequently, when designing identification experiments special attention needs to be given to sensor (actuator) placements and identifiability problems~\cite{haber2017state}. We want to ensure that the system dynamics is sufficiently excited both in the time and in the spatial domain~\cite{haber2014subspace,haber2013moving}

 To briefly summarize, temperature dynamics identification problems belong to a class of Multiple Input Multiple Output (MIMO) identification problems with a data set consisting of a  small number of time and spatial data samples. The classical identification techniques, such as for example, prediction error methods~\cite{ljung1998system}, might prove to be impractical for the estimation of the temperature dynamics. This is mainly due to the infinite-dimensional nature and MIMO structure of the dynamics. A more suitable option is to use subspace identification techniques~\cite{verhaegen2007filtering,van2012subspace,haber2012identification}.

Motivated by these challenges, in this manuscript, we consider the identification and validation problems for temperature dynamics. We have developed an experimental setup consisting of a long aluminum bar whose temperature dynamics is influenced by spatially distributed heat actuators and whose temperature is sensed by spatially distributed thermocouples. We address the noise reduction problem and perform step response and nonlinearity analyses. We combine Predictor-Based Subspace IDentification (PBSID) methods~\cite{houtzager2009varmax,chiuso2007role,houtzager2012recursive} with time series analysis methods~\cite{lutkepohl2005new} to identify  and validate a MIMO state-space model of the experimental setup. We also identify a Kalman innovation state-space form that is used as an observer for the experimental setup. We provide a detailed treatment of model structure selection, validation, and residual analysis problems under different modeling and prediction scenarios. Besides this, we investigate the effect of the sampling period duration on the dynamics of the identified model.  Codes and data-sets used to generate the experimental results are publicly available online\footnote{https://github.com/AleksandarHaber/Machine-Learning-and-Identification-of-Temperture-Dynamics}.

 The experimental setup and preliminary identification results are first presented in our conference publication~\cite{haber2019temperature}. However, except for the experimental setup description, this paper contains novel results. Namely,  this paper contains novel identification results that are obtained by identifying a Kalman innovation state-space model of the experimental setup. Furthermore, in contrast to the conference publication, in this manuscript, we present a detailed procedure for estimating the model order of an AutoRegressive with eXogenous inputs model and a detailed residual analysis of the identified model.
 
 This paper is organized as follows. In Section~\ref{experimentalSetupDescription} we describe the experimental setup and perform the step response and system nonlinearity analyses. In Section~\ref{subspaceIdentification} we describe the identification and model order selection methods as well as the residual analysis. Finally, in Section~\ref{identificationResults} we present the identification results and in Section~\ref{conclusionsSection} we present the conclusions. 
  
\section{Experimental Setup Description, Noise Reduction, Step Response Analysis, and System Nonlinearities}
\label{experimentalSetupDescription}

In this section, we briefly describe the experimental setup. Furthermore, we describe digital filters that were used for the reduction of the measurement noise. Finally, we analyze the step response and the system nonlinearities. The results presented in this section are used in the next section to postulate the system model. This section is based on the preliminary results presented in the conference manuscript~\cite{haber2019temperature}. 

\subsection{Experimental Setup}

\begin{figure}[H]
\centering 
\includegraphics[scale=0.22,trim=0mm 0mm 0mm 0mm ,clip=true]{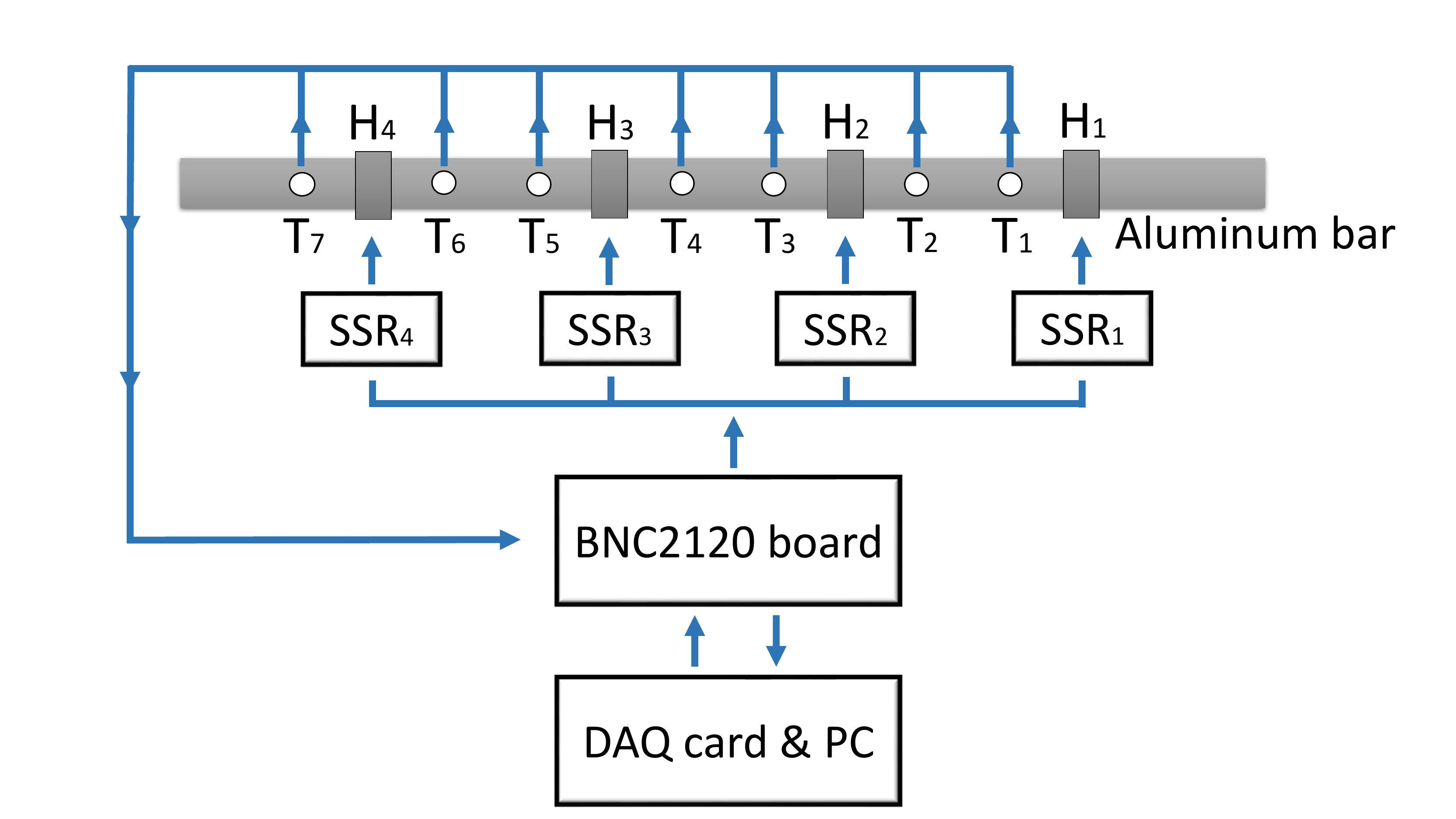}
\caption{The working principle of the experimental setup. The temperature of the long cylindrical bar is sensed and controlled by seven thermocouples ($T_{i}, \; i=1,2,\ldots, 7$) and four band heaters ($H_{i},\; i=1,\ldots, 4$), respectively. The heaters are controlled by four Solid State Relays ($SSR_{i}, \; i=1,\ldots,4$). The SSRs and thermocouples are interfaced with the computer (PC) through a BNC2120 board and a Data AcQuisition (DAQ) card.}
\label{fig:Graph01}
\end{figure}

\begin{figure}[H]
\centering 
\includegraphics[scale=0.26,trim=0mm 0mm 0mm 0mm ,clip=true]{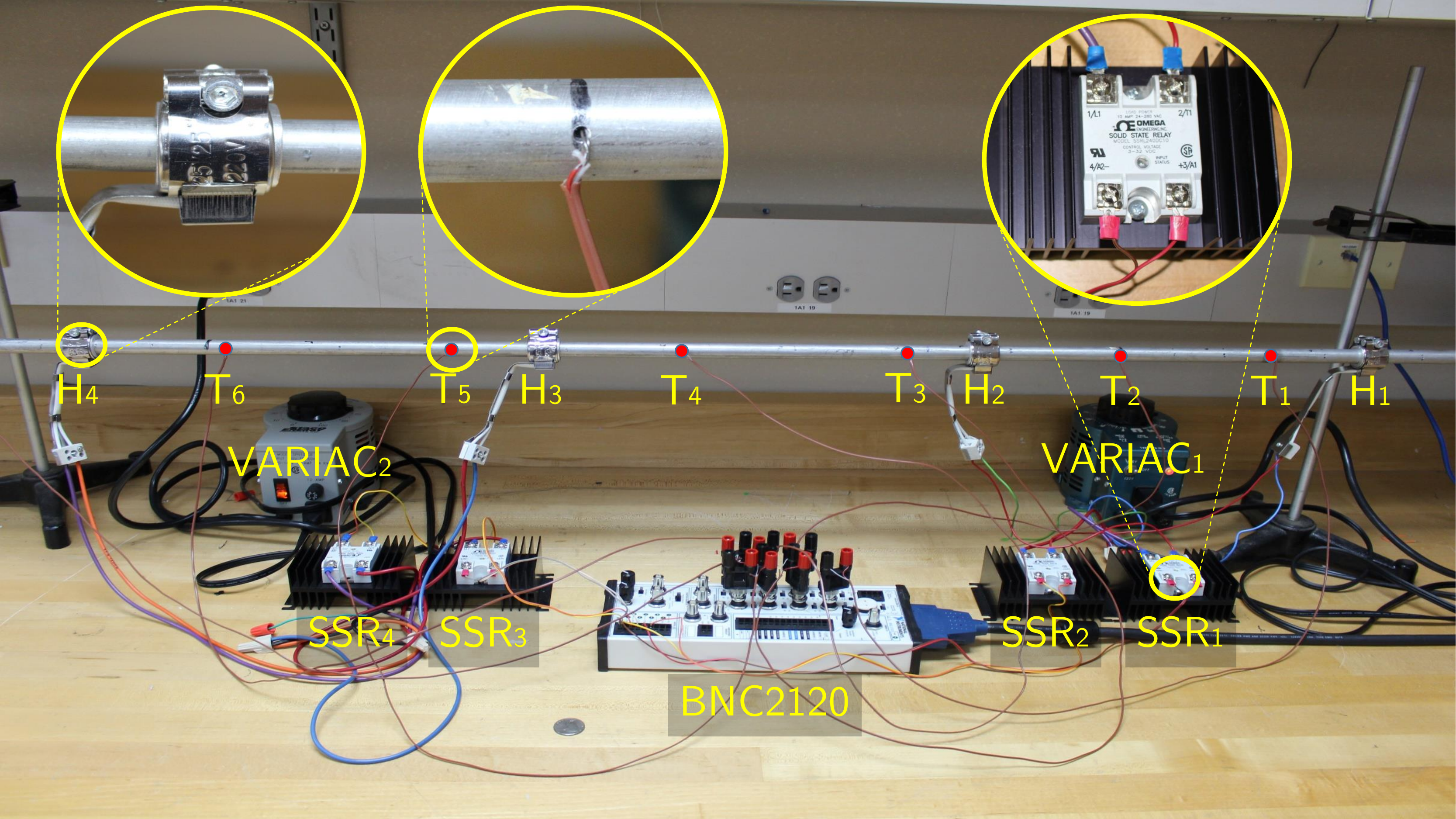}
\caption{The experimental setup photograph. The abbreviations of the components correspond to the abbreviations in Fig.~\ref{fig:Graph01}.}
\label{fig:Graph02}
\end{figure}

\begin{figure}[H]
\centering 
\includegraphics[scale=0.35,trim=0mm 0mm 0mm 0mm ,clip=true]{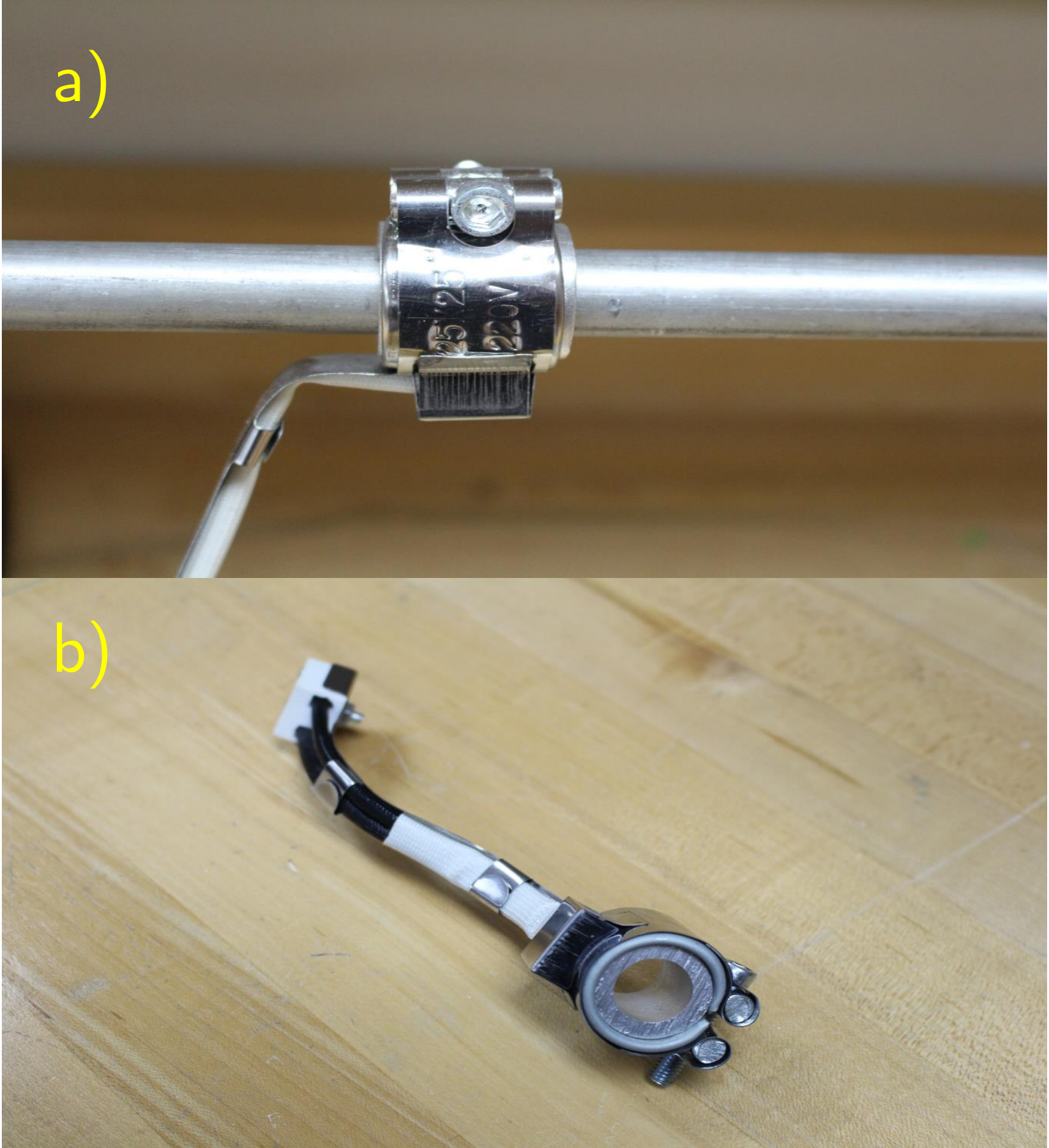}
\caption{(a) The band heater after being mounted on the aluminum bar and (b) the band heater with an aluminum mounting ring.}
\label{fig:Graph03}
\end{figure}

The main components of the experimental setup are shown in Fig.~\ref{fig:Graph01}. A photograph of the setup is shown in Fig.~\ref{fig:Graph02}. The setup consists of a thin cylindrical aluminum bar with the length of $2\; [m]$ and the diameter of $0.015 \; [m]$. We use four spatially distributed band heaters to control the temperature of the bar. The band heaters are denoted by  $H_{i}$, $i=1,2,3,$ and $4$. The rated heater power is $300\; [W]$ and the heater width is $0.03$ $[m]$. Heater photographs are given in Fig.~\ref{fig:Graph03}. The heaters are mounted at the following spatial positions: $0.25,0.65,1.05,$ and $1.55$ $[m]$, measured from the left end of the bar.  Solid State Relays (SSRs) are used to control the heaters. The SSR are denoted  by SSR$_{i}$, $i=1,2,3,$ and $4$, in text and figures. The SSR has three ports. The first port is connected to the digital output ($0-5 \; [V]$) of the BNC2120 data acquisition board. This port is used to receive a low-voltage control signal. The second port is attached to an AC variable autotransformer (VARIAC). The VARIAC provides $0-140$ $[VAC]$ output with the rated current of $10$ $[A]$. The third port is attached to the heater. When the control voltage is high (in the range of $5 \; [V]$), the VARIAC supplies the power to the heater.

We use seven spatially distributed J-type thermocouples to measure the temperature. They are denoted by $T_{i}$, $i=1,2,\ldots, 7$ in text and figures. For the step response analysis, when we only actuate the heater $H_{1}$, the positions of the thermocouples are $0.3,0.35,0.4,0.45,0.5,0.6,$ and $0.7$ $[m]$ (measured from the left end of the bar). In the identification experiment, when all four heaters are actuated, the thermocouple positions are $0.35,0.50,0.70,0.90,1.10,1.50,$ and $1.75$ $[m]$. The thermocouples are placed inside of holes with the depth of $0.0075$ $[m]$ (approximately equal to the bar radius).

To control the setup we use a PC computer with the Windows operating system and the MATLAB programming language. The data analysis and system identification are performed in the MATLAB and Python programming languages. We use the BNC2120 board and a Data AcQuisition (DAQ) card to interface the computer with the thermocouples and SSRs. We use a built-in BNC2120 temperature sensor to measure ambient temperature. This temperature is used for the thermocouple calibration. We use the software cold-junction compensation method~\cite{duff2010two}. The ambient temperature during the performed identification experiments was $26.5$ $[\textdegree{}C]$. The heaters are controller using the Pulse Width Modulation (PWM) technique. 

\begin{figure}[H]
\centering 
\includegraphics[scale=0.18,trim=0mm 0mm 0mm 0mm ,clip=true]{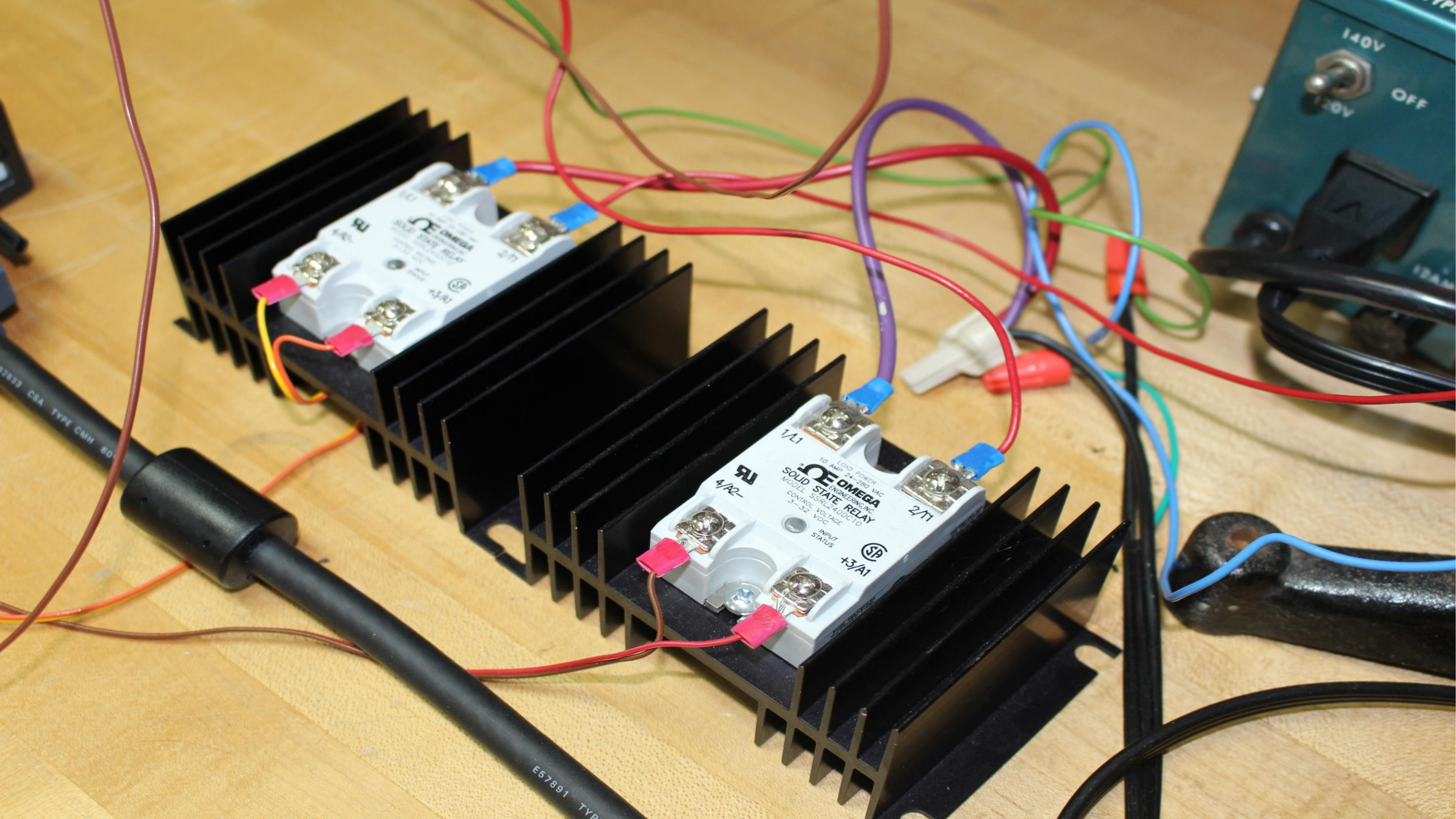}
\caption{Solid State Relays (SSRs) that are  mounted on the heat sinks. }
\label{fig:Graph04}
\end{figure}

\subsection{Data Analysis and Filtering}
In order to successfully identify the system model, we first need to perform a few preparatory steps. If these steps are not performed carefully and correctly, then the identified model can contain modeling errors that might significantly deteriorate the prediction performance. First, we need to understand the origins and the nature of the measurement noise.  Then, we need to design appropriate filters that should attenuate the undesired noise while not affecting the useful signal. Also, we need to eliminate possible outliers in the data.  After these steps are performed, we need to gain crucial insights into the system dynamics. We need to investigate dominant time constants, system bandwidth, system delays, and nonlinearities. Some of these insights can be obtained by performing a step response analysis. The information obtained from these steps is used to properly select the identification sampling frequency, identification control inputs, and the model structure. We first address the noise reduction problem.

\subsubsection{Measurement noise filtering}
\begin{figure}[H]
\centering 
\includegraphics[scale=0.46,trim=0mm 0mm 0mm 0mm ,clip=true]{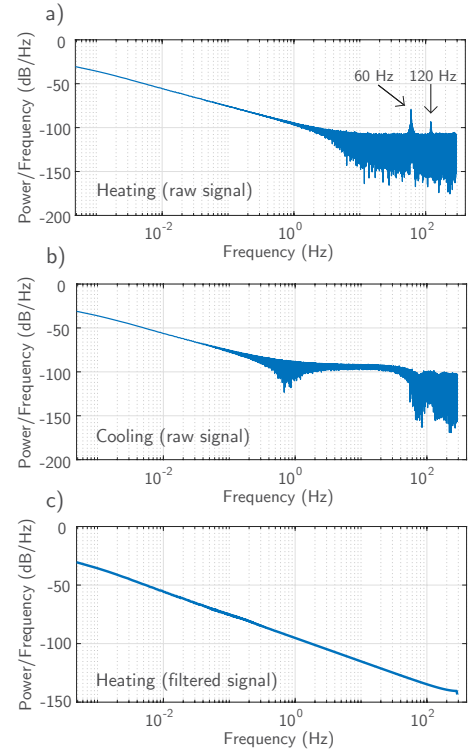}
\caption{PSDs of the temperature signals observed by the thermocouple $T_{1}$. (a) The PSD of the unfiltered signal during the heating process (when the heater $H_{1}$ is active). (b) The PSD of the unfiltered signal during the cooling process (when the heater $H_{1}$ is inactive). (c) The PSD of the filtered signal during the heating process. The signal is filtered using the fourth-order Butterworth filter with the cutoff frequency of $0.2 \; [Hz]$.}
\label{fig:Graph05}
\end{figure}

In order to investigate the properties of the measurement noise, we set the heater $H_{1}$  voltage to $0.9 \; v_{max}$ ($v_{max}=140$ $[VAC]$). In this experiment, the heater $H_{1}$ is positioned at $0.25$ $[m]$. The temperature is sensed by the thermocouple $T_{1}$ located at the position of the heater $H_{1}$. The temperature is recorded during both the heating and cooling processes (when the heater $H_{1}$ is turned off). The time periods of the heating and cooling processes were $2100\; [s]$. We detrend the data and compute the signal Power Spectral Density (PSD). The sampling frequency used to collect the data was $577 \; [Hz]$. The PSDs of the measured signals during the heating and cooling processes are shown in Fig.~\ref{fig:Graph05}(a)  and Fig.~\ref{fig:Graph05}(b), respectively.  In Fig.~\ref{fig:Graph05}(a), we can observe peaks at $60$ $[Hz]$ and $120$ $[Hz]$. Since these peaks are not present during the cooling process in Fig.~\ref{fig:Graph05}(b), we conclude that they originate from the supply voltage provided by the VARIAC. The VARIAC is connected to the $60 \; [Hz]$ power grid. To eliminate these and other high-frequency noise components, we design a fourth-order digital Butterworth filter with the cutoff frequency of $0.2 \; [Hz]$. We can safely select such a low cutoff frequency since the dynamics of the heating process is very slow (this is analyzed in the sequel). Finally, Fig.~\ref{fig:Graph05}(c) shows the PSD of the filtered signal during the heating of the aluminum bar. We can observe that the high-frequency noise components are significantly attenuated.

\subsubsection{Step response analysis} 
\begin{figure}[H]
\centering 
\includegraphics[scale=0.43,trim=0mm 0mm 0mm 0mm ,clip=true]{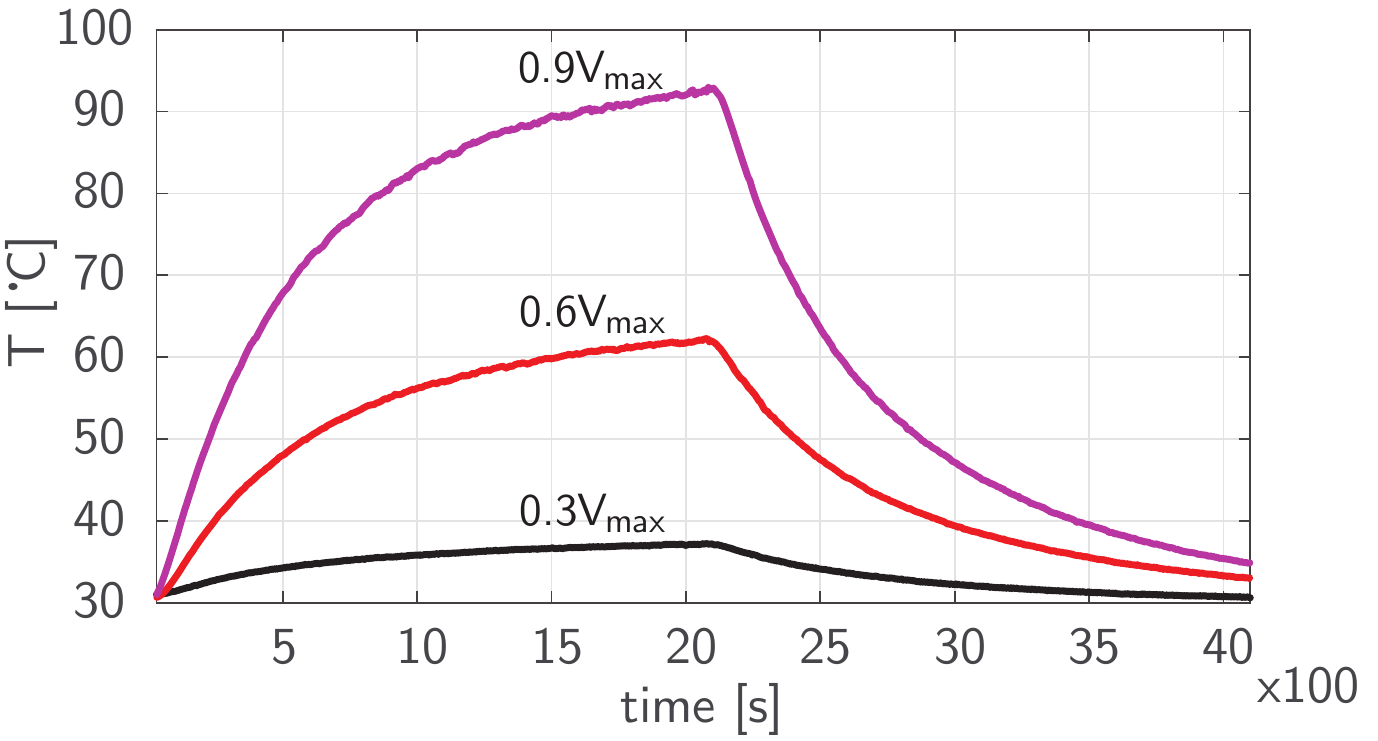}
\caption{The system step and cooling responses observed by the thermocouple $T_{1}$. The step responses are recorded while the voltage was applied to the heater $H_{1}$.}
\label{fig:Graph06}
\end{figure}

We perform the step response analysis for several values of the input voltages. The heater $H_{1}$ is actuated with the voltages of $0.3\; v_{\text{max}}$, $0.6\; v_{\text{max}}$, and $0.9\; v_{\text{max}}$, where $v_{\text{max}}=140$ $[VAC]$ is the maximal voltage generated by the VARIAC. We observe both the heating and the cooling time responses by the thermocouple $T_{1}$. The heating and cooling time periods are both equal to $2100$ $[s]$. We apply the Butterworth filter and eliminate the outlier data.  The filtered responses are shown in Fig.~\ref{fig:Graph06}. We use the recorded time responses to estimate the dominant system time constant. For the heating process, a rough estimate of the system time constant is $750$ $[s]$. This estimate is computed under the assumption that the system response resembles the response of a first-order system. Although in reality the system dynamics is governed by an infinite-dimensional heat equation with linear convection and nonlinear boundary conditions, this simplifying assumption is still useful for drawing important conclusions on the system dynamics. 

To investigate the system linearities/nonlinearities we use a time-domain test described in~\cite{haber1985nonlinearity}. We denote the system response to a zero input by $y_{0}\in \mathbb{R}$. This response is constant and equal to the ambient temperature. Furthermore, we denote the system step responses to the voltages $v_{a}(t)=a s(t)$ and $v_{b}(t)=b s(t)$ by $y_{a}(t) \in \mathbb{R}$ and $y_{b}(t) \in \mathbb{R}$, respectively,  where $s(t) \in \mathbb{R}$ is a step function, $t$  is time, and $a,b\in \mathbb{R}$ are the scaling constants. The index $w_{h}(t) \in \mathbb{R}$, used to evaluate the system nonlinearities, is defined by~\cite{haber1985nonlinearity}:

\begin{align}
w_{h}(t)=\frac{y_{b}(t)-y_{0}}{y_{a}(t)-y_{0}}.
\label{wIndexInputs}
\end{align}

\begin{figure}[H]
\centering 
\includegraphics[scale=0.44,trim=0mm 0mm 0mm 0mm ,clip=true]{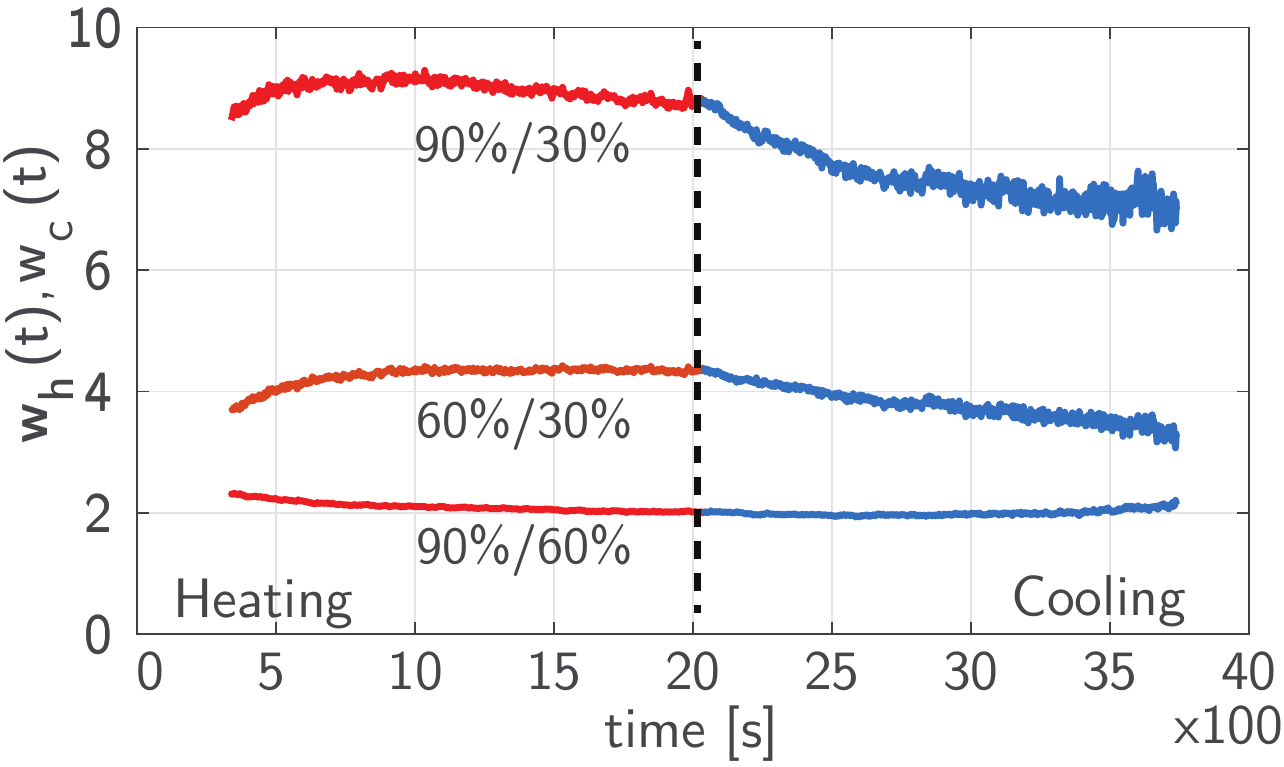}
\caption{The coefficients $w_{h}(t)$ and  $w_{c}(t)$ constants defined in \eqref{wIndexInputs} and \eqref{wIndexStates}, respectively.  The notation $X \% / Y  \%$ denotes the coefficient $w_{h}(t)$ computed for control voltages equal to $X$ and $Y$ percentages of $v_{\text{max}}$.}
\label{fig:Graph07}
\end{figure}

If the system is linear, then we have $w_{h}(t)=b/a$. The system nonlinerities can be detected by the deviation of $w_{h}(t)$ from this ideal constant. For the cooling processes, the index $w_{c}(t) \in \mathbb{R}$ is defined by
\begin{align}
w_{c}(t)=\frac{y_{x_{02}}(t)-y_{0}}{y_{x_{01}}(t)-y_{0}},
\label{wIndexStates}
\end{align}
where $y_{x_{02}}(t) \in \mathbb{R}$ and $y_{x_{01}}(t) \in \mathbb{R}$ are the system outputs for the initial conditions $x_{02}\in \mathbb{R}$ and $x_{01}\in \mathbb{R}$, respectively.  
 Figure~\ref{fig:Graph07} shows the parameters $w_{h}$ and $w_{c}$ that are calculated on the basis of the measurement data. We use the notation $X \% / Y \%$ to denote the coefficient $w_{h}(t)$ computed for $u_{b}= b s(t) $ and $u_{a} = a s(t)$, where $b=(X/100) v_{max}$ and $a=(Y/100) v_{max}$.  That is, the coefficients are computed for $X$ and $Y$ percentages of the maximal voltage. The average values of the coefficient  $w_{h}$ for $90\% / 60\%$, $60\% / 30\% $, and $90\% / 30\% $ are $2.101$, $ 4.271$, and $8.962$, respectively. These averages imply that the output depends approximately quadratically on the applied control voltages $v$. This is an experimental confirmation of the fact that the power of the heat actuator is a quadratic function of the applied voltages. This system nonlinearity can be eliminated by defining control inputs as squares of the applied voltages. These important insights on the system dynamics, nonlinearities, and time constants, that are presented in this section, are used in the next two sections to postulate the model structure and to design an identification experiment.

\section{Subspace Identification, Order Estimation, and Residual Analysis}
\label{subspaceIdentification}

In this section, on the basis of the analysis presented in the previous section, we first postulate the system model. Then, we present an identification algorithm that is a combination of the Predictor Based Subspace IDentification (PBSID) method~\cite{houtzager2009varmax} and the Akaike information criterion method for estimating the order of a Vector AutoRegressive with eXogenous (VARX) terms model. Furthermore, we present a method for state order estimation, and at the end of the section, we briefly summarize a residual analysis method that is originally used for the validation of estimated time-series~\cite{lutkepohl2005new}. The methods presented in this section are used in the next section to identify and validate the system model. The identification algorithm is summarized in Algorithm~\ref{algorithm1} at the end of this section.

\subsection{Notation and Problem Formulation}
To describe the subspace identification algorithm, it is first necessary to introduce the following notation. Let $\mathbf{w}\in \mathbb{R}^{p}$ be an arbitrary real vector and let $q$ and $r$ be non-negative integers ($q\le  r$). The vector $\mathbf{w}_{q,r}\in \mathbb{R}^{r-q+1}$, which we refer to as the \textit{lifted vector}, is defined by
\begin{align}
\mathbf{w}_{q,r}=\begin{bmatrix}\mathbf{w}_{q}^{T} & \mathbf{w}_{q+1}^{T} & \ldots & \mathbf{w}_{r}^{T} \end{bmatrix}^{T}.
\label{liftedVector}
\end{align}
With the vector $\mathbf{w}_{q,r}$, we associate a matrix $W_{q,r}^{(l)}\in \mathbb{R}^{(r-q+1)\times (l+1)}$, where $l$ is a non-negative integer. The matrix $W_{q,r}^{(l)}$ is referred to as the \textit{data matrix} and is defined as follows
\begin{align}
W_{q,r}^{(l)}=\begin{bmatrix}\mathbf{w}_{q,r} & \mathbf{w}_{q+1,r+1} & \ldots & \mathbf{w}_{q+l,r+l} \end{bmatrix}.
\label{dataMatrixDefinition}
\end{align}
Let $X$ be a matrix of arbitrary dimensions. The notation $X(:, d_{1}:d_{2})$ denotes a matrix constructed from the columns $d_{1},d_{1}+1,\ldots, d_{2}$ of the matrix $X$. Similarly, the notation $X( d_{1}:d_{2},:)$ denotes a matrix constructed from the rows $d_{1},d_{1}+1,\ldots, d_{2}$ of the matrix $X$. This is a standard MATLAB notation for selecting the matrix columns (rows) that is often used in the system identification literature~\cite{verhaegen2007filtering}.

Consider the following state-space model:
\begin{align}
\mathbf{d}_{k+1} & =A\mathbf{d}_{k}+B\mathbf{u}_{k}+\mathbf{w}_{k}, \label{stateEquation1original} \\
 \mathbf{y}_{k} & =C\mathbf{d}_{k}+\mathbf{g}_{k}, \label{outputEquation1original} 
\end{align}
 where $k$ is a discrete-time instant, $\mathbf{y}_{k}\in \mathbb{R}^{7}$ is the vector whose entries are temperature measurements by $7$ thermocouples, $\mathbf{u}_{k}\in \mathbb{R}^{4}$ is the control input whose entries are squares of control voltages,  $\mathbf{d}_{k}\in \mathbb{R}^{n}$ is the state, $\mathbf{w}_{k}\in \mathbb{R}^{n}$ is the disturbance vector, and $\mathbf{q}_{k}\in \mathbb{R}^{7}$ is the measurement noise, $A\in \mathbb{R}^{n\times n}$, $B \in \mathbb{R}^{n\times 4}$,  and $C\in \mathbb{R}^{7\times n}$ are the system matrices, and $n$ is the state order.
With the model \eqref{stateEquation1original} and \eqref{outputEquation1original}, we can associate a Kalman innovation state-space model~\cite{verhaegen2007filtering}:
\begin{align}
 \mathbf{x}_{k+1} & =A\mathbf{x}_{k}+B\mathbf{u}_{k}+K\mathbf{e}_{k}, \label{stateEquation1} \\
 \mathbf{y}_{k} & =C\mathbf{x}_{k}+\mathbf{e}_{k}, \label{outputEquation1} 
\end{align}
where  $\mathbf{e}_{k}\in \mathbb{R}^{7}$ is the innovation process,  $K \in \mathbb{R}^{n\times 7}$ is the Kalman gain matrix, and $\mathbf{x}_{k}\in \mathbb{R}^{n}$ is the state. It should be noted that the innovation representation \eqref{stateEquation1}-\eqref{outputEquation1} is an observer for the state-space model \eqref{stateEquation1original}-\eqref{outputEquation1original}. The identification problem can be formulated as follows
\begin{framed}
\textbf{Identification problem.} Using the set of the input-output data $\mathcal{I}=\Big\{ (\mathbf{u}_{k},\mathbf{y}_{k}) \; | \; k=0,1,2,\ldots, N  \Big\}$ estimate the state order $n$, system matrices $A$, $B$, $C$, and $K$, and the initial states of the state-space models \eqref{stateEquation1original}-\eqref{outputEquation1original} and \eqref{stateEquation1}-\eqref{outputEquation1}.
\end{framed}
 By substituting \eqref{outputEquation1} in \eqref{stateEquation1}, we obtain the following state-space model
\begin{align}
 \mathbf{x}_{k+1} & =\tilde{A}\mathbf{x}_{k}+\tilde{B}\mathbf{z}_{k}, \label{stateEquation2} \\
 \mathbf{y}_{k} & =C\mathbf{x}_{k}+\mathbf{e}_{k}, \label{outputEquation2} 
\end{align}
where 
\begin{align}
\tilde{A}=A-KC,\; \tilde{B}=\begin{bmatrix}B & K \end{bmatrix},\; \mathbf{z}_{k}=\begin{bmatrix}\mathbf{u}_{k} \\ \mathbf{y}_{k}  \end{bmatrix},  
\label{predictorExplanation}
\end{align}
and  $\tilde{A}\in \mathbb{R}^{n\times n}$, $\tilde{B}\in \mathbb{R}^{n\times 11}$, and $\mathbf{z}_{k} \in \mathbb{R}^{11}$.
For two positive integers $f$ and $p$, $f\le p$, we introduce the following matrices:
\begin{align}
L_{p-1}=\begin{bmatrix}\tilde{A}^{p-1}\tilde{B} & \tilde{A}^{p-2}\tilde{B} & \ldots & \tilde{A}\tilde{B} & \tilde{B} \end{bmatrix}, \notag \\
O_{f-1}=\begin{bmatrix} \big(C \big)^{T} & \big(C\tilde{A}  \big)^{T} & \ldots & \big(C\tilde{A}^{f-1} \big)^{T}  \end{bmatrix}^{T},
\label{liftedMatrices}
\end{align}
where $K_{p-1}\in \mathbb{R}^{n\times 11p}$ and $O_{f-1}\in \mathbb{R}^{7 f \times n}$. The parameters $f$ and $p$ are referred to as the \textit{future} and \textit{past} windows, respectively. The subspace identification method consists of two estimation steps. The first step is the estimation of the VARX model. The second step consists of the estimation of the state sequence and system matrices.

\subsection{Estimation of the VARX model}
From \eqref{stateEquation1}, we have:
\begin{align}
\mathbf{x}_{k}= \tilde{A}^{p}\mathbf{x}_{k-p}+L_{p-1}\mathbf{z}_{k-p,k-1},
\label{state1}
\end{align}
where the matrix $L_{p-1}$ is defined in \eqref{liftedMatrices} and the vector $\mathbf{z}_{k-p,k-1}\in \mathbb{R}^{p}$ is defined using the notation introduced in \eqref{liftedVector}. A standard assumption in subspace identification is that
\begin{align}
\tilde{A}^{s}\approx 0, \;\; \forall s \ge p.
\label{assumptionSubspace}
\end{align}
The physical justification of this assumption follows from the following facts. All the eigenvalues of the matrix $\tilde{A}$ are inside of the unit circle. Due to this, we can select the past window $p$ such that \eqref{assumptionSubspace} is satisfied. Taking this assumption into account, we have:
\begin{align}
\mathbf{x}_{k}\approx L_{p-1}\mathbf{z}_{k-p,k-1}.
\label{stateExpressed}
\end{align}
From \eqref{outputEquation2} and \eqref{stateExpressed}, we have
\begin{align}
\mathbf{y}_{k}\approx M_{p-1} \mathbf{z}_{k-p,k-1}+\mathbf{e}_{k},
\label{predictor1}
\end{align}  
where the matrix $M_{p-1}\in \mathbb{R}^{7\times 11 p}$ is the matrix of Markov parameters, defined by
\begin{align}
M_{p-1}=CL_{p-1}=\begin{bmatrix} C\tilde{A}^{p-1}\tilde{B} & C\tilde{A}^{p-2}\tilde{B} & \ldots & C\tilde{A}\tilde{B} & C\tilde{B}  \end{bmatrix}.
\label{markovMatrix}
\end{align}
The equation \eqref{predictor1} is a one step ahead predictor for the system output and it has a VARX model form. On the basis of this equation, we estimate the matrix $M_{p-1}$ as the solution of the following least-squares problem
\begin{align}
&\min_{M_{p-1}}  \left\| E_{p-1}  \right\|_{F}^{2}, \label{lsProblem} \\
& E_{p-1}= Y_{p,p}^{(l)}-M_{p-1}Z_{0,p-1}^{(l)}, \label{lsExplanation}
\end{align}
where the matrices $Y_{p,p}^{(l)} \in \mathbb{R}^{7\times (l+1)}$ and $Z_{0,p-1}^{(l)}\in \mathbb{R}^{11p\times (l+1)}$ are defined using the notation in \eqref{dataMatrixDefinition}. The solution is given by
\begin{align}
\hat{M}_{p-1}=Y_{p,p}^{(l)}\big(Z_{0,p-1}^{(l)}\big)^{T} \Big(  Z_{0,p-1}^{(l)} \big(Z_{0,p-1}^{(l)}\big)^{T} \Big)^{-1}.
\label{solutionLS}
\end{align}
In \eqref{solutionLS} we assumed that the data matrix $Z_{0,p-1}^{(l)}$ has a full row rank. This is, we assume that the \textit{persistency of excitation condition} is satisfied~\cite{verhaegen2007filtering}.

\subsubsection{VARX Model Order Selection}

 An important problem in estimation theory is the selection of the past window $p$ of the VARX model~\eqref{predictor1}. We use the Akaike Information Criterion (AIC) for the estimation of multiple time series~\cite{lutkepohl2005new} to estimate the parameter $p$. Namely, we choose the parameter $p$ that minimizes the following scalar quantity: 
\begin{align}
\text{AIC}(p)= \ln{\det{\hat{\Gamma}_{p-1}}} + \frac{2}{l+1}\big( 77p\big), \;\; p=1,2,\ldots, p_{\text{max}},
\label{Akaike1}
\end{align}
where
\begin{align}
\hat{\Gamma}_{p-1}=\frac{1}{l+1} \hat{E}_{p-1}\hat{E}_{p-1}^{T}, \;\; \hat{E}_{p-1}= Y_{p,p}^{(l)}-\hat{M}_{p-1}Z_{0,p-1}^{(l)}.
\end{align}
A few comments are in order. The matrix $\hat{\Gamma}_{p-1}$ is the maximum likelihood estimate of the residual covariance matrix. In \eqref{Akaike1}, the first term takes into account the model prediction accuracy while the second term penalizes the model complexity. The scalar quantity $77w$ denotes the total number of estimated parameters (the total number of entries of the matrix $\hat{M}_{p-1}$). The parameter $p_{max}$ is the maximal past window  that needs to be tested. This parameter is a user choice. To summarize, to find the estimate of the past window, we need to compute $p_{\text{max}}$ least-squares solutions \eqref{solutionLS}, and using such solutions to find $w$ for which the quantity in \eqref{Akaike1} has the smallest value. 

\subsection{State Sequence and System Matrices Estimation}

By multiplying \eqref{stateExpressed} by $O_{f-1}$, defined in \eqref{liftedMatrices}, and taking into account \eqref{assumptionSubspace}, we obtain:
\begin{align}
O_{f-1}\mathbf{x}_{k}\approx  Q_{p-1} \mathbf{z}_{k-p,k-1},
\label{finalExpression}
\end{align}
where 
\begin{align}
Q_{p-1}= \begin{bmatrix} M_{p-1}    \\  \big[ 0_{7 \times 11 } \;\;  M_{p-1}\big(:, 1: 11(p-1)   \big) \big] \\   \vdots \\  \big[ 0_{7 \times 11(f-1) } \;\;  M_{p-1}\big(:, 1: 11(p-f+1)   \big) \big] \end{bmatrix},
\label{Qmatrix}
\end{align}
and $0_{d_{1} \times d_{2} }$ is a $d_{1}\times d_{2}$ matrix of zeros and $Q_{p-1}\in \mathbb{R}^{7f\times 11 p}$. Let  the matrix $\hat{Q}_{p-1}$ denote the estimate of the matrix $Q_{p-1}$ constructed using the estimated matrix $\hat{M}_{p-1}$.  From \eqref{finalExpression}, we have
\begin{align}
O_{f-1}X_{p,p}^{(l)}\approx \hat{Q}_{p-1} Z_{0,p-1}^{(l)},
\label{data1}
\end{align}
where $X_{p,p}^{(l)} \in \mathbb{R}^{n\times (l+1)}$ is the state sequence. The last equation tells us that we can estimate the state sequence by estimating  the row space of the matrix $\hat{Q}_{p-1} Z_{k-p,k-1}^{(l)}$. This can be achieved as follows. Compute the Singular Value Decomposition (SVD) of the matrix $\hat{Q}_{p-1} Z_{0,p-1}^{(l)}$
\begin{align}
\hat{Q}_{p-1} Z_{0,p-1}^{(l)}=U\Sigma V^{T},
\label{svdDecomposition}
\end{align}
where $U\in \mathbb{R}^{7f\times 7f}$, $\Sigma\in \mathbb{R}^{7f \times (l+1)}$, and $V\in \mathbb{R}^{(l+1)\times (l+1)}$. Let $\hat{n}$ be an estimate of the state order. The estimate of the state sequence can be computed as follows
\begin{align}
\hat{X}_{p,p}^{(l)}=\Sigma(1:\hat{n},1:\hat{n})^{1/2} V_{T}(1:\hat{n},:),
\label{stateSequenceEstimate}
\end{align}
where $V_{T}=V^{T}$ and $\hat{X}_{p,p}^{(l)}\in \mathbb{R}^{\hat{n}\times (l+1)}$. It follows that
\begin{align}
\hat{X}_{p,p}^{(l)}=\begin{bmatrix}\hat{\mathbf{x}}_{p} & \hat{\mathbf{x}}_{p+1} & \ldots & \hat{\mathbf{x}}_{p+l}   \end{bmatrix}.
\label{estimatedStateDecomposed}
\end{align}
Using the estimated state sequence, and equations  \eqref{stateEquation2} and \eqref{outputEquation2}, we have:
\begin{align}
& \hat{X}_{p+1,p+1}^{(l-1)}\approx Q S, \label{FinalEquations1} \\
& Y_{p,p}^{(l)}\approx C \hat{X}_{p,p}^{(l)}+E_{p,p}^{(l)},  \label{FinalEquations2}
\end{align}
where
\begin{align}
Q=\begin{bmatrix} \tilde{A} & \tilde{B} \end{bmatrix}, S=\begin{bmatrix} \hat{X}_{p,p}^{(l-1)} \\ Z_{p,p}^{(l-1)}  \end{bmatrix},
\label{definitionFinal}
\end{align}
and the matrices $\hat{X}_{p+1,p+1}^{(l-1)}\in \mathbb{R}^{\hat{n}\times l}$ and  $\hat{X}_{p,p}^{(l-1)}\in  \mathbb{R}^{\hat{n}\times l} $ are formed from the estimated sequence $\{ \hat{\mathbf{x}}_{k} \}$ using the definition \eqref{dataMatrixDefinition}.  Similarly, the matrices $Y_{p,p}^{(l)}\in \mathbb{R}^{7 \times (l+1)}$ and  $E_{p,p}^{(l)}\in \mathbb{R}^{7 \times (l+1)}$ are formed from the vectors $\mathbf{y}_{k}$ and $\mathbf{e}_{k}$, respectively,  using the definition \eqref{dataMatrixDefinition}. The system matrices $\tilde{A}$, $\tilde{B}$, $K$, and $C$ are estimated by solving the following least-squares problems
\begin{align}
\min_{Q} \left\|  \hat{X}_{p+1,p+1}^{(l-1)} - Q S \right\|_{F}^{2}, \;\; \min_{C} \left\|Y_{p,p}^{(l)}-  C \hat{X}_{p,p}^{(l)}  \right\|_{F}^{2}.
\label{finalEstimates}
\end{align}
The solutions of the problems in \eqref{finalEstimates} are given by
\begin{align}
\hat{Q}= \hat{X}_{p+1,p+1}^{(l-1)}S^{T}\big(SS^{T} \big)^{-1}, \;  \hat{C}=Y_{p,p}^{(l)}\big(\hat{X}_{p,p}^{(l)} \big)^{T}\Big(\hat{X}_{p,p}^{(l)} \big(\hat{X}_{p,p}^{(l)} \big)^{T} \Big)^{-1}.
\label{solutionLS2}
\end{align}
The system matrices are determined by 
\begin{align}
& \hat{\tilde{A}}=\hat{Q}(:,1:\hat{n}), \hat{B}=\hat{Q}(:,\hat{n}+1:\hat{n}+4), \notag \\ 
& \hat{K}=\hat{Q}(:,\hat{n}+5:\hat{n}+11),\;\; \hat{A}=\hat{\tilde{A}}+\hat{K}\hat{C}.
\label{systemMatrices}
\end{align}

\subsubsection{Model Selection}
\label{modelSelection}

Similarly to the problem of estimating the past window of the VARX model~\eqref{predictor1}, in order to implement the second step, we need to estimate the future window and the model order. Every pair of state orders and future windows uniquely  determines a system model. In the sequel, we describe the procedure for model selection. The set $\mathcal{S}$, defined by 
\begin{small}
\begin{align}
\mathcal{S}=\big\{(i,j) \; | \; i=1,\ldots, n_{\text{max}}, \; j=f_{\text{min}}(i),f_{\text{min}}(i)+1,\ldots, f_{\text{max}}  \big\},
\label{setS}
\end{align}
\end{small}
denotes the set of ordered pairs of state orders and future windows. The positive integers $n_{\text{max}}$, $f_{\text{min}}(i)$, and $f_{\text{max}}$ denote the maximal state order, minimal future window (that depends on $i$), and the maximal future window, respectively. The maximal state order is a user choice, and in our numerical experiments $n_{\text{max}}=40$. The minimal future window depends on the state order $i$ since we need to ensure that the observability matrix $O_{j-1}$, defined by substituting $f$ by $j$ in \eqref{liftedMatrices}, has more rows that columns. That is $f_{\text{min}}(i)=\ceil[\big]{i/7}$, where $\ceil[\big]{x}$ denotes a ceiling function of the argument $x$. The maximal future window is $f_{\text{max}}=\hat{p}$, where $\hat{p}$ is the past window estimate. For every element in the set $\mathcal{S}$, we estimate the state sequence and the system matrices. That is, for every pair of state orders and future windows,  we estimate a system model.  The ''optimal'' model is then selected by testing the prediction performance of the estimated models using a new set of data (that was not used for the identification of the VARX model and the state matrices). This set of data is referred to as the $\textit{validation data}$ and is defined as  follows $\mathcal{V}=\Big\{ (\underline{\mathbf{u}}_{k},\underline{\mathbf{y}}_{k}) \; | \; k=0,1,2,\ldots, N_{1}  \Big\}$, where the underline notation is used to distinguish validation from identification data sets. The validation data is first used to estimate an initial state. Depending on the model type and prediction scenario, we distinguish three different cases for estimating the model order.

\begin{itemize}
\item \textit{Method A: Model selection on the basis of the open-loop simulation of \eqref{stateEquation1original}-\eqref{outputEquation1original}.} We estimate the initial state of the state-space model \eqref{stateEquation1original}-\eqref{outputEquation1original} by solving the following least-squares problem
\begin{align}
\min_{\underline{\mathbf{d}}_{0}} \left\|  \underline{\mathbf{y}}_{0,h-1}  - \hat{D}_{h-1} \underline{\mathbf{u}}_{0,h-1} - \hat{O}_{h-1}\underline{\mathbf{d}}_{0} \right\|_{2}^{2},
\label{LSproblemState}
\end{align}
where $h\ll N_{1}$,  is the length of the sequence used for the initial state estimation and $\underline{\mathbf{d}}_{0}\in \mathbb{R}^{n} $  is an unknown initial state.  The matrix $\hat{O}_{h-1}$ is the observability matrix formed using the estimates $\hat{A}$ and $\hat{C}$, and 
\begin{align}
\hat{D}_{h-1}=\begin{bmatrix} 0 & 0 & 0 &\ldots & 0  \\ \hat{C}\hat{B} & 0 & 0 & \ldots & 0 \\  \hat{C}\hat{A}\hat{B} & \hat{C}\hat{B} & 0 & \ldots &0 \\  \vdots & \vdots & \ddots &  & \vdots  \\   \hat{C}\hat{A}^{h-2}\hat{B} & \hat{C}\hat{A}^{h-3}\hat{B} & \ldots & \hat{C}\hat{B} &0   \end{bmatrix}. \label{vectorsIntroduced}
\end{align}
The solution of the least-squares problem \eqref{LSproblemState} is given by
\begin{align}
\underline{\hat{\mathbf{d}}}_{0}=\hat{O}^{\dagger}_{h-1}\big( \underline{\mathbf{y}}_{0,h-1}-\hat{D}_{h-1}\underline{\mathbf{u}}_{0,h-1}  \big),
\label{initialStateEstimate}
\end{align}
where $\hat{O}^{\dagger}_{h-1}\in \mathbb{R}^{n\times 7b}$ is the pseudo-inverse of $\hat{O}_{h-1}$. The estimated initial state, estimated system matrices, and the validation input sequence $\big\{\underline{\mathbf{u}}_{k}\;  | \; k=0,1,\ldots, N_{1}  \big\}$  are used to simulate the known part of the system dynamics \eqref{stateEquation1original}-\eqref{outputEquation1original} (we do not know the disturbance vector $\mathbf{w}_{k}$ and the measurement noise vector $\mathbf{g}_{k}$). In this way, we generate an estimate of the validation output sequence $\big\{ \hat{\underline{\mathbf{y}}}_{k} \; |\; k=0,1,2,\ldots, N_{1} \big\}$. The relative error between this sequence and the measured validation sequence is computed as follows
\begin{align}
e=\left\|\underline{\mathbf{y}}_{0,N_{1}}- \hat{\underline{\mathbf{y}}}_{0,N_{1}}  \right\|_{2}/ \left\| \underline{\mathbf{y}}_{0,N_{1}} \right\|_{2}.
\label{validationErrors}
\end{align}
where the vectors $\underline{\mathbf{y}}_{0,N_{1}}$ and $ \hat{\underline{\mathbf{y}}}_{0,N_{1}}$ are defined in accordance with the definition~\eqref{liftedVector}. We compute the model error \eqref{validationErrors} for all the pairs of state orders and future windows in the set $\mathcal{S}$. We select the pair for which $e$ has the smallest value. \textit{This model selection method is referred to as Method A.}

\item \textit{Method B: Model selection on the basis of the open-loop simulation of \eqref{stateEquation2}-\eqref{outputEquation2}.} This method is based on the simulation of the Kalman innovation representation~\eqref{stateEquation2}-\eqref{outputEquation2}. Similarly to the Method A, using the state-space model \eqref{stateEquation2}-\eqref{outputEquation2} we form a least-squares problem and estimate the initial state. This least-squares problem is similar to \eqref{LSproblemState},  except the fact that $\underline{\mathbf{d}}_{0}$, $\hat{A}$, $\hat{B}$, and $\underline{\mathbf{u}}_{k}$ are formally replaced by $\underline{\mathbf{x}}_{0}$, $\hat{\tilde{A}}$, $\hat{\tilde{B}}=\begin{bmatrix}\hat{B} & \hat{K}  \end{bmatrix}$, and $\underline{\mathbf{z}}_{k}=\begin{bmatrix} \underline{\mathbf{u}}_{k}^{T} & \underline{\mathbf{y}}_{k}^{T}   \end{bmatrix}^{T}$. Once the initial state is estimated, the state equation \eqref{stateEquation2} is simulated starting from $\underline{\mathbf{x}}_{0}$ and $\underline{\mathbf{z}}_{0}$. In the simulation steps $k=1,2,\ldots,N_{1}$, we use the vector $\hat{\underline{\mathbf{z}}}_{k}=\begin{bmatrix} \underline{\mathbf{u}}_{k}^{T} & \hat{\underline{\mathbf{y}}}_{k}^{T} \end{bmatrix}^{T}$, where $ \hat{\underline{\mathbf{y}}}_{k} =C\hat{\underline{\mathbf{x}}}_{k}$. That is, we use a simulated output at the time instant $k$ to compute the state-sequence $\hat{\underline{\mathbf{x}}}_{k+1}$, $k=1,2,\ldots, N_{1}$. Using this data, we compute the error~\eqref{validationErrors}. We select the model order and the future window for which this error has the smallest value.  \textit{This model selection method is referred to as Method B.}

\item \textit{Method C: Model selection on the basis of the closed-loop simulation of \eqref{stateEquation2}-\eqref{outputEquation2}.} The initial state is estimated using the procedure used to estimate the initial state in Method B. However, in Method C, the simulated output sequence is generated by simulating the model \eqref{stateEquation2}-\eqref{outputEquation2} using the vector $\underline{\mathbf{z}}_{k}=\begin{bmatrix} \underline{\mathbf{u}}_{k}^{T} & \underline{\mathbf{y}}_{k}^{T} \end{bmatrix}^{T}$. That is, we use the measured validation data sequence to simulate the model. We select the model order and the future window that produce the smallest value of the error \eqref{validationErrors}. \textit{This model selection method is referred to as Method C.}
\end{itemize}

A few comments are in order. Instead of using the error \eqref{validationErrors} to validate the model, we can use  the Variance Accounted For (VAF)~\cite{verhaegen2007filtering}. This parameter is computed by using the validation output and the simulated output. The VAF of $100\%$ corresponds to a perfect match between the outputs of the identified and real systems, for more details see~\cite{verhaegen2007filtering}. The state order and the future window are selected such that the VAF has the maximal value. Our results show that the VAF achieves the maximal value for the parameters that  produce the smallest value of the validation error. That is, these two measures can be used interchangeably.

\subsection{Residual test}
More information about the identified model quality can be obtained by analyzing the residuals. Let the error sequence between the measured output and the simulated output be defined as follows
\begin{align}
\boldsymbol{\epsilon}_{k}= \underline{\mathbf{y}}_{k}-\hat{\underline{\mathbf{y}}}_{k}, \;\; k=0,1,2,\ldots, N_{1}.
\label{errorSequence}
\end{align}
This sequence is referred to as the \textit{residual sequence}. Ideally, if the model captures all the information about the system, then the residual sequence is uncorrelated and it should have white-noise properties. In practice, the residual sequence will most likely be correlated due to the model errors. Despite this, by computing the autocorrelation matrices and performing a white-noise hypothesis test on such matrices, we can see how close is the residual sequence to the ideal case. The residual analysis used in this manuscript  is based on the residual analysis for multiple time series presented in~\cite{lutkepohl2005new}.

First, we estimate the autocovariance matrices as follows
\begin{align}
\hat{\Delta}_{i} = \frac{1}{N_{1}} \sum_{k=i}^{N_{1}} \big( \boldsymbol{\epsilon}_{k} -\bar{\boldsymbol{\epsilon}}   \big) \big(  \boldsymbol{\epsilon}_{k-i} -\bar{\boldsymbol{\epsilon}}     \big)^{T} , \;\; i=0,1,\ldots, N_{1}-l_{1},
\label{autoCovarianceMatrices}
\end{align}
where $l_{1}$ is a positive integer (selected by the user) and  $\bar{\boldsymbol{\epsilon}}$ is the mean residual defined by 
\begin{align}
\bar{\boldsymbol{\epsilon}}=\frac{1}{N_{1}}\sum_{k=0}^{N_{1}}  \boldsymbol{\epsilon}_{k}.
\label{meanVector}
\end{align}
The estimates of the autocorrelation matrices are defined by
\begin{align}
\hat{\Gamma}_{i}=\hat{D}^{-1}_{0}\hat{\Delta}_{i}\hat{D}^{-1}_{0},
\label{estimatedAutoCorrelation}
\end{align}
where $\hat{D}_{0}$ is a diagonal matrix with the main diagonal entries equal to the square roots of the diagonal entries of $\hat{\Delta}_{0}$. Let  $\hat{\gamma}_{b,s}(i)$ denote the $(b,s)th$ entry of the matrix $\hat{\Gamma}_{i}$. To visualize the correlations, we plot  $\hat{\gamma}_{b,s}(i)$ for different values of the lag $i$. If the values of $\hat{\gamma}_{b,s}(i)$ are large, then there is a significant correlation between the $b$th entry of $\boldsymbol{\epsilon}_{k}$ and the $s$th entry of $\boldsymbol{\epsilon}_{k-i}$. If the value of $\hat{\gamma}_{b,s}(i)$ is small,  then most likely there is no correlation between the corresponding entries. The words ''small'' and ''large'' should be interpreted with respect to statistical significance levels of the white noise sequence. This is explained in the sequel. 

Let $\gamma_{b,s}(i)$ denote the $(b,s)th$ entry of the ''true'' autocorrelation matrix, denoted by $\Gamma_{i}$. We want to test the hypothesis that the process $\boldsymbol{\epsilon}_{k}$ is a white noise. For a white noise process, $\gamma_{b,s}(i) =0$, for $i>0$. In~\cite{lutkepohl2005new} the following white-noise hypothesis test is proposed:
\begin{align}
H_{0}: \; \gamma_{b,s}(i) =0,\;\; \text{against} \;\;  H_{1}:\; \gamma_{b,s}(i) \ne 0,
\label{hypothesisTesting}
\end{align}
where $H_{0}$ is the null hypothesis, that is, the hypothesis that the sequence is a white noise, and $H_{1}$ is the alternative hypothesis. At the $5\%$ level, the null hypothesis is rejected if $|\hat{\gamma}_{b,s}(i)|>2/\sqrt{N_{1}}$.

The identification algorithm is summarized in Algorithm~\ref{algorithm1}.

\begin{algorithm}
  \caption{Subspace Identification Algorithm}
  \begin{algorithmic}[1]
    \Statex \textbullet~\textbf{Input data:} Identification and validation data sets.
    \Statex \textbullet~\textbf{User parameters:} Maximal past window $p_{\text{max}}\in \mathbb{Z}^{+}$, window $h$ for computing the initial state \eqref{initialStateEstimate}, and the parameters $n_{\text{max}}\in \mathbb{Z}^{+}$ and  $f_{\text{max}}\in \mathbb{Z}^{+}$ defining the set $\mathcal{S}$ in \eqref{setS}.
    \State For $p=1,2,\ldots,p_{max}$, estimate the VARX models using \eqref{solutionLS} and select the VARX model order (the value of $\hat{p}$) by minimizing  \eqref{Akaike1}. This step is performed using the identification data set.
    \State For all the pairs of state orders and future windows belonging to the set $\mathcal{S}$, compute the state sequence \eqref{stateSequenceEstimate} and estimate the system matrices \eqref{systemMatrices}. This step is performed using the identification data set.
    \State If Method A is used for model selection, for every estimated system in step 2, estimate the initial state using \eqref{initialStateEstimate} and simulate the system \eqref{stateEquation1original}-\eqref{outputEquation1original} using the validation input sequence. Select the final model for which the validation error $e$, defined in \eqref{validationErrors}, has the smallest value. For model selection methods B and C perform equivalent steps described in Section~\ref{modelSelection}.
    \end{algorithmic}
\label{algorithm1}
\end{algorithm}

\section{Identification Results} 
\label{identificationResults}
In this section, we first address the problems of selecting identification control inputs and sampling frequency. Then, we present the results of identifying the VARX model. Finally, we present the identification and validation results.

\subsection{Input Sequence and Sampling Frequency }

First, we choose the identification input sequence. Control voltages are selected as uniformly distributed random scalars belonging to the interval $[ 0,1 ]$. Such control voltages produce control inputs $\mathbf{u}_{k}$ that jointly with the generated outputs $\mathbf{y}_{k}$ produce the data matrix  $Z_{0,p-1}^{(l)}$ of full rank. That is, this choice of control voltages ensures that the persistence of excitation condition is fully satisfied~\cite{verhaegen2007filtering}. We generate identification and validation input sequences. These sequences together with the corresponding output data are referred to as the \textit{identification and validation data sequences}. The lengths of the identification and validation sequences are $180$ and $120$ discrete-time samples, respectively. The control inputs are converted to PWM signals that are used to control the SSRs, for more details see Section~\ref{experimentalSetupDescription}. 

Next, we address the problem of selecting the sampling frequency (sampling period). Due to the fact that the temperature dynamics is relatively slow, special attention needs to be paid to the selection of the identification sampling frequency. First, the raw data is sampled with the maximal sampling frequency that is limited by the data acquisition hardware and software. The average value of the sampling frequency is approximately $577$ $[Hz]$. Secondly, we apply a low-pass filter and downsample the raw data to a lower frequency that is used to identify and validate the model.  The problem of choosing this down-sampled frequency is a critical one since a poor choice can lead to a model that is not useful from the practical point of view. The sampling frequency should be approximately ten times larger than the bandwidth of interest~\cite{verhaegen2007filtering}. We choose the bandwidth of interest as the system bandwidth. On the basis of the step response analysis, in Section~\ref{experimentalSetupDescription} we estimated the system bandwidth. Our estimate is the interval $[0, 1/750]$ $[\text{rad}/ s]$, where $750$ is an estimate of the system time constant. This estimate is obtained under a first-order order approximation of the system step response. From this estimate, it follows that the optimal value of the identification sampling period should be $440$ $[s]$. However, this sampling period is not practically feasible due to the measurement time constraints. Namely, to successfully implement the identification algorithm we need a relatively large number of data samples. \textit{The sampling period of $440$ $[s]$ would result in the identification experiment that would last for several days.} In practice, we are not able to continuously measure the system response for more than several hours. Due to this, we decrease the sampling frequency such that the identification and validation data can be collected in a reasonable time interval.
For comparison, we choose two sampling frequencies, and consequently, we identify two systems:
\begin{enumerate}
\item \textit{System 1}. The sampling period and the period of the PWM signal are $96$ $[s]$. 
\item \textit{System 2}. The sampling period and the period of the PWM signal are  $208$ $[s]$.
\end{enumerate}
While analyzing the identification data, it should be kept in mind that since our sampling frequencies are higher than the recommended one, we expect that some of the eigenvalues of the identified system matrix $A$ will be close to the unit circle~\cite{verhaegen2007filtering}.

\subsection{VARX Model Estimation}

\begin{figure}[H]
\centering 
\includegraphics[scale=0.31,trim=0mm 0mm 0mm 0mm ,clip=true]{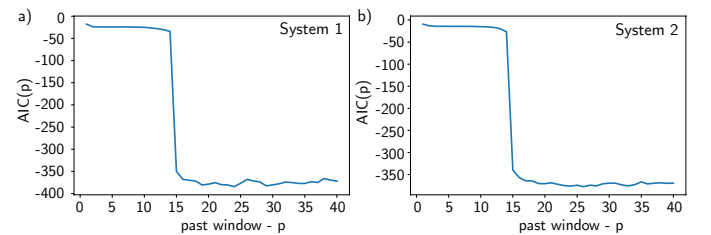}
\caption{AIC graphs for estimating the past window. (a) System 1. (b) System 2.}
\label{fig:Graph08}
\end{figure}

First, we estimate the VARX models and the past windows. The AIC values as functions of past windows are shown in Fig.~\ref{fig:Graph08}. For both cases, we observe steep declines of the AIC values around the past window value of  $15$. The optimal past window estimates are $\hat{p}=24$ and $\hat{p}=26$ for System 1 and System 2, respectively. 

\subsection{State Estimation and Validation}

\begin{figure}[H]
\centering 
\includegraphics[scale=0.31,trim=0mm 0mm 0mm 0mm ,clip=true]{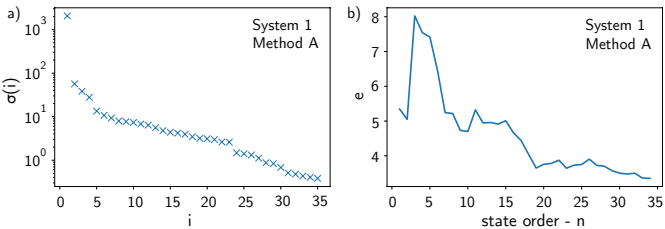}
\caption{(a) Singular values of the matrix $\hat{Q}_{p-1} Z_{0,p-1}^{(l)}$, see equation~\eqref{svdDecomposition}.  (b) Relative identification error, defined in \eqref{validationErrors}, as a function of the state order $n$. The results are generated for System 1, Method A, and $p=24$ and $f=5$.}
\label{fig:Graph09}
\end{figure}

For the estimated past window of $\hat{p}=24$, we obtain the estimates of the future window of $\hat{f}=24$ and the state order of $\hat{n}=34$. Figure~\ref{fig:Graph09} illustrates the identification performance for other values of state orders. Figure~\ref{fig:Graph09}(a) shows the singular values of the matrix $\hat{Q}_{p-1} Z_{0,p-1}^{(l)}$, see equation~\eqref{svdDecomposition}. Figure~\ref{fig:Graph09}(b) shows the relative error defined in \eqref{validationErrors} as a function of the state order. The results shown in Fig.~\ref{fig:Graph09}(b)  imply that the order selection on the basis of the singular value gaps in Fig.~\ref{fig:Graph09}(a) can be misleading. For example, in Fig.~\ref{fig:Graph09}(a), we can observe a gap between the first few singular values and the remaining ones. This can lead us to the conclusion that the optimal state order is equal to $1$ or $4$. However, from Fig.~\ref{fig:Graph09}(b) we see that for the state order of $4$, the relative error achieves its maximum. While interpreting these results it should be kept in mind that singular values are computed on the basis of the identification data set, whereas the relative error is computed for the validation data set. 

\begin{figure}[H]
\centering 
\includegraphics[scale=0.29,trim=0mm 0mm 0mm 0mm ,clip=true]{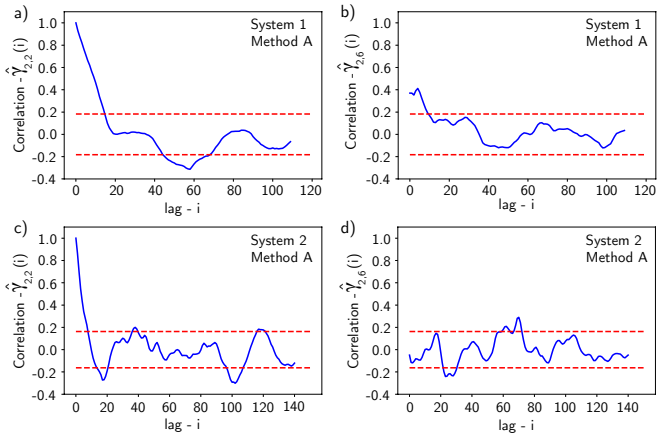}
\caption{The residual analysis results.  (a) and (b)  correlation coefficients for System 1. (c) and (d) correlation coefficients for System 2. The dashed lines represent $\pm 2/\sqrt{N_{1}}$ bounds for white-noise hypothesis testing. }
\label{fig:Graph10}
\end{figure}

Figure \eqref{fig:Graph10} shows the residual test results. The results are obtained for the parameters giving the smallest value of the relative error. For System 1, the results are obtained for $\hat{p}=24$, $\hat{f}=5$, and $\hat{n}=34$. For System 2, the results are obtained for  $\hat{p}=26$, $\hat{f}=18$, and $\hat{n}=13$. The smallest values of the relative error for System 1 and System 2 are $e=3.34 \%$ and $e= 4.73 \%$, respectively. For both systems, we can observe that there is a strong correlation between $\hat{\gamma}_{2,2}$ for short values of the lag $i$. This indicates that there is still some useful information in the data that is not captured by the identified model. On the other hand, this correlation is not so significant for $\hat{\gamma}_{2,6}$.

\begin{figure}[H]
\centering 
\includegraphics[scale=0.29,trim=0mm 0mm 0mm 0mm ,clip=true]{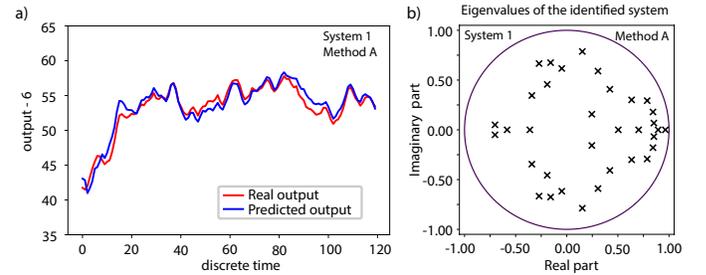}
\caption{ (a) Prediction performance of the identified model of System 1. (b)  Eigenvalues of the identified model of System 1. The results are generated  the parameters: $\hat{p}=24$, $\hat{f}=5$, and $\hat{n}=34$.}
\label{fig:Graph11}
\end{figure}

\begin{figure}[H]
\centering 
\includegraphics[scale=0.29,trim=0mm 0mm 0mm 0mm ,clip=true]{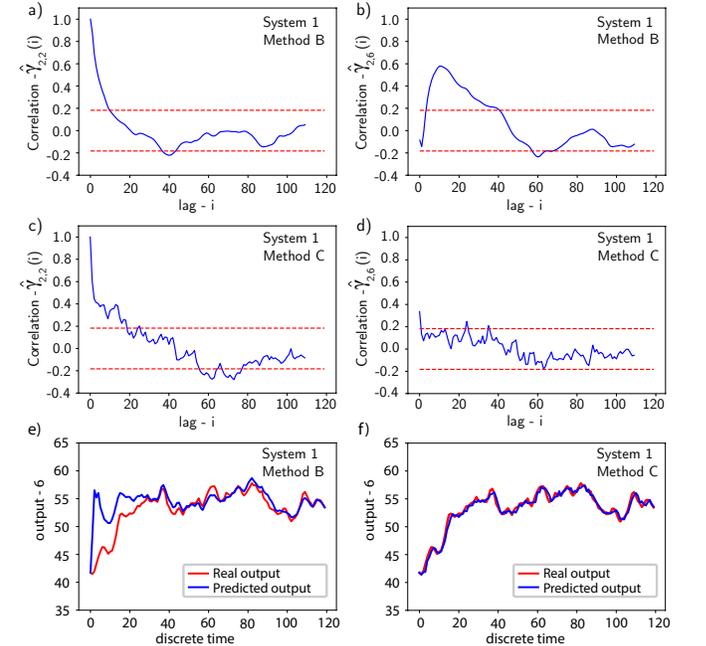}
\caption{(a) and (b)  correlation coefficients for System 1 whose model is selected using Method B. (c) and (d)  correlation coefficients for System 1 whose model is selected using Method C. Prediction performance of System 1 whose model is selected using (e) Method B and (f) Method C. }
\label{fig:Graph12}
\end{figure}

Figure~\ref{fig:Graph11} shows the prediction performance of the identified model of System 1. The model is selected using Method A. Figure~\ref{fig:Graph11}(b) shows the eigenvalues of the identified matrix $A$ of System 1. It can be observed that the identified model is stable. Furthermore, some of the eigenvalues are close to the unit circle which is probably due to the fact that the identification sampling frequency has a higher value than the optimal one. Figure~\ref{fig:Graph12} shows the residual analysis of the models of System 1 selected using Methods B and C, as well as the corresponding prediction performances. For Method B, the final model relative error is $ 5.83 \%$ that is obtained for $\hat{n}=12$, $\hat{f}=11$. For Method C, the final model relative error is $1.18 \%$ that is obtained for $\hat{n}=6$, $\hat{f}=1$. 

It can be observed that model selection Method C produces the best prediction performance. This is due to the fact that it uses real validation measurements to predict the state. Furthermore, the model selected using Method C has the smallest state order. \textit{That is, if we can perform the state prediction on the basis of the real data, then we do not need a model of a high order.}



\section{Conclusion}
\label{conclusionsSection}
In this manuscript, we considered the problem of data-driven modeling of temperature dynamics. Namely, we used subspace identification and time series analysis methods to identify and validate a model of a heat conduction experimental setup. The experimental setup consists of a long aluminum bar whose temperature is controlled by spatially distributed heat actuators. The temperature of the bar is sensed by distributed thermocouples. We addressed the noise reduction problem and performed step response and nonlinearity analyses. We provided detailed treatments of model structure selection, validation, and residual analysis problems. We demonstrated that the temperature dynamics can be relatively accurately represented by a Kalman innovation form of a linear state-space model. We demonstrated that the optimal model order depends on the simulation scenario performed in the model validation step. The lowest model order and best prediction performance is achieved when the simulation is performed in closed-loop.

\section*{Acknowledgment}
This work was supported by the PSC-CUNY Award A (61303-00 49). The author would like to thank Thomas Rodberg, Melvin Summerville, Francesco Pecora, and Mobin Chowdhury for technical help. 
\ifCLASSOPTIONcaptionsoff
  \newpage
\fi



\bibliographystyle{IEEEtran}
%
\bibliography{bare_jrnl}

%




\end{document}